\documentclass[%
 reprint,
superscriptaddress,
 amsmath,amssymb,
 aps,
pra,
]{revtex4-2}

\usepackage[pdftex]{graphicx}
\usepackage{dcolumn}
\usepackage{physics}
\usepackage{bm}
\usepackage{xcolor}
\usepackage{tikz}

\providecommand{\V}[1]{\boldsymbol{#1}}
\providecommand{\M}[1]{\mathbf{#1}}
\providecommand{\T}[1]{\mathrm{#1}}

\providecommand{\J}{\T{i}}

\begin{document}

\preprint{APS/123-QED}

\title{Fundamental Bounds on the Polarizability of Macroscopic Scatterers}

\author{Lukas Jelinek}
\affiliation{Department of Electromagnetic Field, Czech Technical University in Prague}
\email{lukas.jelinek@fel.cvut.cz}

\author{Miloslav Capek}
\affiliation{Department of Electromagnetic Field, Czech Technical University in Prague}

\date{\today}

\begin{abstract}
Polarizability predicts how an object responds to an incident electromagnetic field, the interactions between small particles, or the optical forces exerted upon them. Polarizability is responsible for the effective-medium properties of artificial materials or metasurfaces. Despite significant progress in all these areas, it is unclear what the limits of the strength of such interactions are or, more specifically, what the upper bounds on the polarizability of a given spatial region that an unknown and designed particle would occupy are. This work connects the electromagnetic field description via an integral equation with a dual formulation of quadratic programming to derive fundamental bounds on components of all four polarizability tensors or on their specific combinations. In particular, the work establishes an intuitive visualization of what strong polarizability means and how strong it can be. The developed fundamental bound also answers which materials and domains are best for the given demands on polarizability. These findings establish a versatile platform that can accommodate a wide range of demands on polarizable bodies, providing an absolute measure of their performance against which the results of human-powered or automated design procedures can be compared. 
\end{abstract}

\maketitle


\section{Introduction}
Polarizability is an essential characteristic of electrically small atomic, molecular, or macroscopic scatterers, upholding the optical properties of matter~\cite{2000_Hohm_Vacuum}, characterizing the optical forces~\cite{Gordon1973,Ashkin1978} and explaining London's dipole-dipole interactions~\cite{London1930,Hermann2017}. In macroscopic electrodynamics, polarizability gives rise to the properties of metasurfaces~\cite{2020_Iyer_TAP} and artificial materials~\cite{1994_Lindell_book,2017_Monticone_RPP}. Overall, polarizability is of paramount importance across electromagnetics.

Despite its enormous impact, the fundamental bounds on polarization strength are not formulated. Attempts have been made in the case of microscopic objects~\cite{Tang1969,Burrows1976}, but their scope, being the estimates of true polarizabilities of atoms and molecules, is different. No account has yet been made of the possible spatial modification of the scatterer. In the macroscopic domain, polarizability tensors have been successfully employed in setting shape-independent bounds on antenna metrics~\cite{Gustafsson2007,
Gustafsson2010} or setting the sum rules of transmission via opaque screens~\cite{Gustafsson2009}. Here, however, the intention is the opposite: to set bounds on the polarizabilities themselves.

The purpose of this paper is to develop computational fundamental bounds that, for a given material, electric size, and coordinate axes, give the maximum allowed dipolar response of a prescribed design region. The bounds are formulated building on the success of fundamental bounds based on quadratic programming~\cite{Nocedal2006} in other electromagnetic metrics~\cite{Gustafsson2013,Gustafsson2015,Jelinek2017,Chao2022}. In particular, the introduction of the complex power constraint~\cite{Gustafsson2020} is a key step that enables the current formulation.




\section{Theory Derived From An Example}
\label{sec:Theory}

\begin{figure*}
    \centering
    \includegraphics[width=\linewidth]{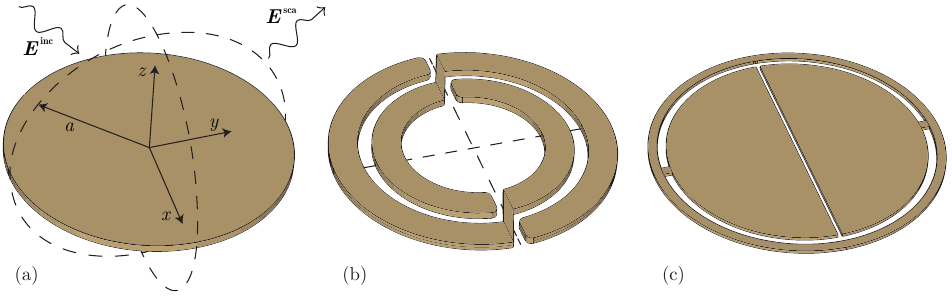}
    \caption{(a) A disc of external radius~$a$, which is also the radius of its smallest circumscribing sphere. The disc is excited by an incident field, producing a scattered field. At electrically small sizes~$ka \ll 1$, the scattered field results dominantly from the dipolar response. (b) A non-bianisotropic split-ring resonator carved into the disc exhibiting resonant magnetic polarizability~$\alpha_{zz}^\T{mm}$. (c) An inversely symmetric electric resonator carved into the disc exhibiting resonant electric polarizability~$\alpha_{yy}^\T{ee}$.}
    \label{fig:FigNBSRR}
\end{figure*}

Assume a highly conducting metallic disc~\footnote{Possible dielectric support of such a structure is left aside for brevity, but can be added with no modification of the theory.} of a given electric size lying in the $xy$ plane in otherwise free space, see Fig.~\ref{fig:FigNBSRR}a. Further assume we would like to carve out the metal to obtain a design of a non-bianisotropic particle with a strong magnetic response in a time-harmonic steady state~\footnote{Time convention~$\T{exp \left(- \J \omega t \right)}$ is assumed.} at angular frequency~$\omega$ (corresponding to wavenumber~$k = \omega / c$, where~$c$ is the speed of light). Inspired by the development of magnetic metamaterials~\cite{Marqus2007},  a non-bianisotropic split ring resonator (NBSRR) \cite{GarcaGarca2005} from Fig.~\ref{fig:FigNBSRR}b tuned close to its resonance could be the scatterer of choice. The question, however, is whether a better design can be found and what the best design looks like.

Two difficulties arise in answering the question. First, one needs to quantify what ``best'' means. Second, removing the material from the disc region in an automated manner to obtain a particle with the best magnetic polarizability is a non-convex optimization problem and there is no guarantee that a global optimum will be found. The second issue is the motivation to formulate a fundamental bound that provides at least an upper limit on the global optimum and, therefore, an estimate of how far human- or computer-aided design can go. The first issue demands a quantification of polarization properties.

To quantify the response of an electrically small particle to the incident electric~$\V{E}^\T{inc}$ and magnetic~$\V{B}^\T{inc}$ field, the polarizability tensors~$\V{\alpha}$~\cite{Tretyakov2003} defined via
\begin{equation}
    \mqty(c \V{p} \\ \V{m}) = \sqrt{\dfrac{\varepsilon}{\mu}} 
    \mqty(\V{\alpha}_\T{ee} & \V{\alpha}_\T{em} \\ \V{\alpha}_\T{me} & \V{\alpha}_\T{mm} ) 
    \mqty (\V{E}^\T{inc}\left( \M{0} \right) \\ c \V{B}^\T{inc} \left( \M{0} \right) )
\label{eq:polarizDef}
\end{equation}
are commonly used, connecting incident fields at the position of the particle (taken as the center of its smallest circumscribing sphere, also coinciding with the coordinate center), and induced electric~$\V{p}$ and magnetic~$\V{m}$ dipole moments. The material parameters of the background medium (permittivity~$\varepsilon$, permeability~$\mu$) are included so that all polarizability tensors have units of volume.

In the case of the aforementioned magnetically polarizable particle, and especially in the case of split ring resonators~\cite{PendryHoldenRobbinsEtAl1999}, we would like to find a particle with polarizability~$\alpha^\T{mm}_{zz}$, see coordinate axes in Fig.~\ref{fig:FigNBSRR}a, having the highest absolute value with both the positive and negative real parts being appealing~\cite{1994_Lindell_book,PendryHoldenRobbinsEtAl1999,2017_Monticone_RPP}. At the same time, we would like to suppress the loss in the particle (ohmic and radiation) as much as possible since these consume the energy of the incident wave without giving any polarization. Lastly, for non-bianisotropic particles, we would like to suppress magneto-electric coupling~\cite{Tretyakov2003}, \textit{i.e.}, polarizability tensors~$\V{\alpha}^\T{em},\V{\alpha}^\T{me}$.

Having the studied region discretized and using the electric field integral equation, see Appendix~\ref{app:A} for details, there are direct ways to evaluate the polarizability of a particular design~\cite{IshimaruLeeKugaEtAl2003,Arango2013,2013_Araque_Metamaterials,2014_Asadchy_PNFA,2016_Yazdi_PIERM,2017_Jelinek_IET}. The six-wave method according to~\cite{2017_Jelinek_IET} is used in this paper. These schemes can directly be used for topology optimization~\cite{Bendse2004,Molesky2018,Capek2023}, though we must remember that these optimization problems are non-convex and there is no guarantee that the best design has been found. In contrast, the fundamental bound must result from convex problems, which ensure global optimality. The price to pay is that there is typically a gap between the fundamental bound and the performance achieved by topology optimizers, and this gap should be as small as possible.

One possible way to formulate the fundamental bound on magnetic polarizability is as follows. When a non-bianisotropic scatterer is to have a high polarization $\alpha^\T{mm}_{zz}$, it means that homogeneous magnetic excitation~$\V{B}^\T{inc} = B_z^\T{inc} \M{z}_0$ should induce a high magnitude of the magnetic moment~$\V{m} = m_z \M{z}_0$, and, at the same time, the induced electric dipole moment~$\V{p}$ should vanish. Since the geometry of the best particle is not known, we use the equivalence principle and represent the reaction of the particle to the incident field by electric current density in free space, assuming that the current can be shaped as desired to reach the optimum performance. This ensures that we obtain an upper bound on polarization, as every design in the same region can, by equivalence, be replaced by a current in free space. 

Without any other constraints, the current density can be made unbounded in amplitude, leading to an unbounded magnetic moment and unbounded polarizability, which is unacceptable. To resolve this issue, we note that induced current density in any real design is bound together with excitation via Maxwell’s equations. Assuming a description via the electric field integral equation and Galerkin’s method from Appendix~\ref{app:A}, Maxwell's equations are represented by a matrix equation
\begin{equation}
    \left( \M{Z}_0 + \M{Z}_\rho \right) \M{I} = \M{V},
    \label{eq:Maxwell}
\end{equation}
where matrix~$\M{Z}_0$ represents free-space interactions, matrix~$\M{Z}_\rho$ represents material distribution, vector~$\M{I}$ represents equivalent current density representing the particle and vector~$\M{V}$ represents excitation. For a given material distribution and a given excitation, the solution to~\eqref{eq:Maxwell} is unique. Similar to~\cite{2020_Gustafsson_NJP}, we can, however, relax this equation to a complex power balance
\begin{equation}
    \M{I}^\T{H} \left( \M{Z}_0 + \M{Z}_\rho \right) \M{I} = \M{I}^\T{H} \M{V} \approx \J \omega \left(  \V{p}^* \cdot \V{E}^\T{inc}\left( \M{0} \right) + \V{m}^* \cdot \V{B}^\T{inc} \left( \M{0} \right)  \right),
    \label{eq:PextX1}
\end{equation}
the real part of which is proportional to the cycle-mean power extinct by the particle described by equivalent current density~$\V{J}$, and where the details of the last approximation are given in Appendix~\ref{app:B}. The superscript~$^\T{H}$ denotes a Hermitian conjugate and~$^*$ denotes a complex conjugate.

Assume now that the system is excited by the magnetic field~$\V{B}^\T{inc} = B_z^\T{inc} \left( x,y \right) \M{z}_0$, which is approximately constant around the origin~$c B_z^\T{inc} \left( \M{0} \right) = 1 \, \T{Vm}^{-1} $, and that the incident electric field approximately vanishes. An example of such excitation is given in Appendix~\ref{app:C}. The power balance gives
\begin{equation}
    \M{I}^\T{H} \M{V} \approx \J \omega m_z^* B^\T{inc}_z = \J \omega \varepsilon \alpha_{zz}^{\T{mm},*} \left| c B_z^\T{inc} \right|^2,
    \label{eq:PextX2}
\end{equation}
which, apart from constants and complex conjugation, measures the desired magnetic polarizability and, importantly, the ratio (assuming a lossless material background)
\begin{equation}
    \dfrac{\T{Re}\left\{ \M{I}^\T{H} \M{V}\right\}}{\T{Im}\left\{ \M{I}^\T{H} \M{V}\right\}} \approx 
    \dfrac{\T{Im}\left\{ m_z \right\}}{\T{Re}\left\{ m_z \right\}} = 
    \dfrac{\T{Im}\left\{ \alpha_{zz}^\T{mm} \right\}}{\T{Re}\left\{ \alpha_{zz}^\T{mm} \right\}}
\end{equation}
quantifies the reactive behavior of the equivalent current density (particle).

With this knowledge, let us state the following optimization problem
\begin{equation}
    \begin{aligned}
\max \limits_\M{I} \quad & \left| m_z \right|^2  \\
  \T{s.t.} \quad     & \M{I}^\T{H} \M{ZI} = \M{I}^\T{H} \M{V} \\
  & \dfrac{ \T{Im} \left\{ \M{I}^\T{H} \M{V} \right\} } { \T{Re} \left\{ \M{I}^\T{H} \M{V} \right\} } = \beta , \,\beta  \in \left( - \infty ,\infty \right) \\
  & \V{p} = \M{0}, \, m_x = 0, \, m_y = 0
    \end{aligned}
    \label{eq:opt-problem}
\end{equation}
where vector~$\M{V}$ is given by the aforementioned magnetic-only excitation. We claim that for a given design region (support of the optimal current density~$\M{I}$), the resulting magnetic moment and, due to the specific excitation, the resulting magnetic polarizability component, represent the strongest dipolar response of all possible designs in the same region, forming an upper bound on polarizability strength. The fundamental bound is parametrized by parameter~$\beta$, giving the reactive behavior of the scatterer.

\subsection{Numerical Results}
\label{sec:IIA}

\begin{figure}
    \centering
    \includegraphics[width=\linewidth]{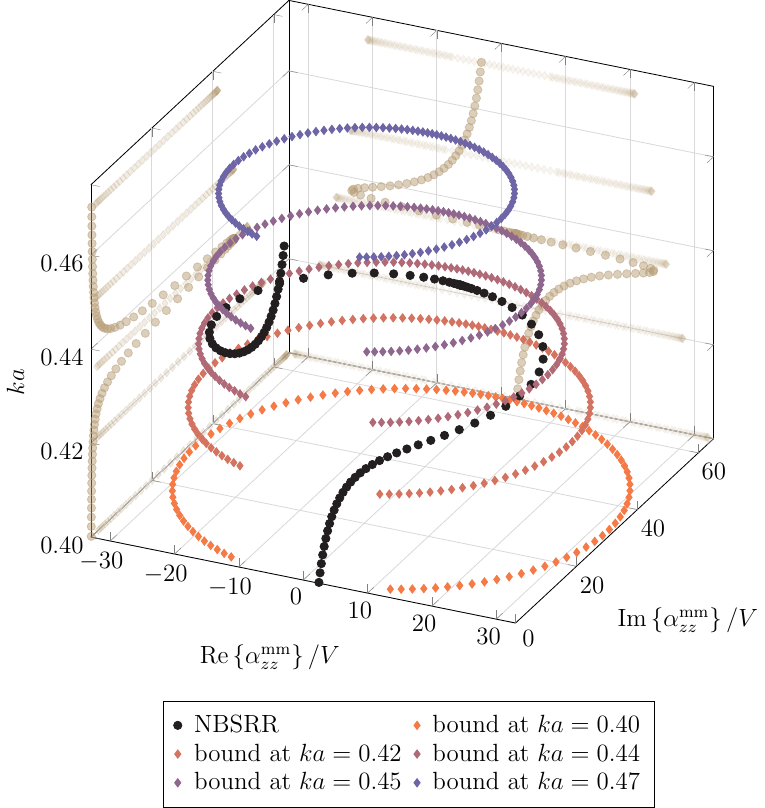}
    \caption{Magnetic polarizability of NBSRR, see Fig.~\ref{fig:FigNBSRR}b, normalized by volume~$V$ of the smallest sphere circumscribing the resonator ($a$ being the radius of such a sphere). The width of the resonator's strip equals~$5a/28$, the width of the slot equals~$a/14$, and the width of the ring split equals~$a/7$. The figure also shows the fundamental bound evaluated via~\eqref{eq:opt-problem} at electric sizes~$ka = \left\{0.40, 0.42, 0.44, 0.45, 0.47 \right\}$. Three-dimensional trajectories are supplemented by their two-dimensional projections (beige) where the narrow-band resonance of the NBSRR is clearly observable. The projections of the bound trajectories delimit the feasible region of the polarizability of any realized design.}
    \label{fig:polarNBSRR_3}
\end{figure}

The magnetic polarizability~$\alpha_{zz}^\T{mm}$ of the NBSRR from Fig.~\ref{fig:FigNBSRR}b, made of a good conductor (the equivalent of copper at frequency~1\,GHz), was retrieved by the six-wave method~\cite{2017_Jelinek_IET} and is depicted by black markers in Fig.~\ref{fig:polarNBSRR_3}. The magnetic excitation from Appendix~\ref{app:C} with~$\V{B}^\T{inc} = B_z^\T{inc} \left(x,y\right) \M{z}_0, \, c B_z^\T{inc} \left( \M{0} \right) \approx 1 \, \T{Vm}^{-1}$ was employed.

Polarizability is plotted as a three-dimensional trajectory in the~$\left[\T{Re} \left\{ \alpha_{zz}^\T{mm} \right\}, \T{Im} \left\{ \alpha_{zz}^\T{mm} \right\}, ka \right]$ space, where the selectric size~$ka$ plays two roles. First, the higher the electric size, the higher the dipolar radiation and, thus, the lower the polarizability potential. Second, it serves as a parameter equivalent to~$\beta$ from~\eqref{eq:opt-problem} sweeping along the trajectory. Along with the polarizability of the NBSRR, five results of optimization problem~\eqref{eq:opt-problem} are also shown. The fundamental bound is evaluated separately at any desired value of electric size, forming a circle parametrized by parameter~$\beta$. When the fundamental bound is evaluated at different electrical sizes, it forms a tube representing a polarizability which cannot be exceeded by any design made of the same material (copper at frequency~1\,GHz) in the same region (disc of radius~$a$). All real designs, represented by three-dimensional trajectories, must lie inside this tube. The fundamental bound also clearly shows that the polarizability potential decays with increasing electric size which is due to radiation loss. 

\begin{figure}
    \centering
    \includegraphics[width=\linewidth]{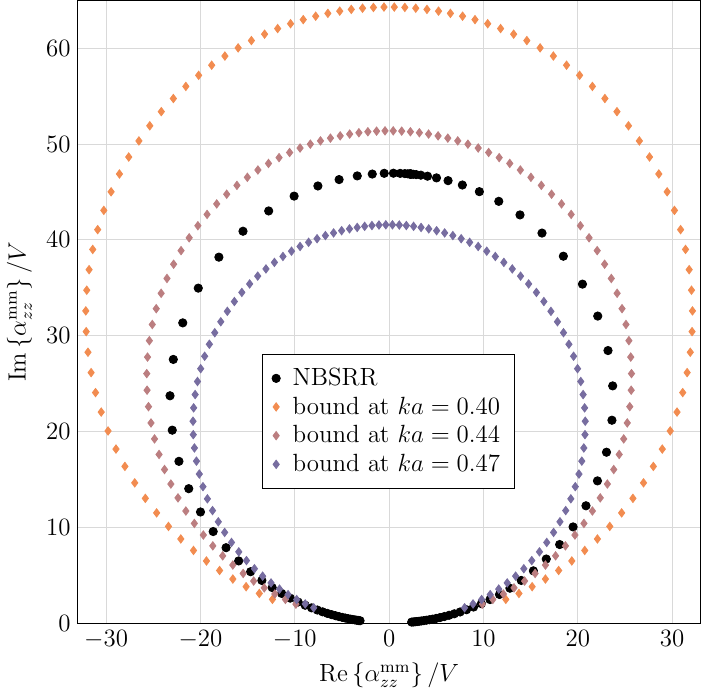}
    \caption{Top projection of the polarizability data from~Fig.~\ref{fig:polarNBSRR_3}. The figure also shows the fundamental bound evaluated via~\eqref{eq:opt-problem} at electric sizes~$ka = \left\{0.40, 0.44, 0.47 \right\}$.}
    \label{fig:polarNBSRR_2}
\end{figure}

The black markers in Fig.~\ref{fig:polarNBSRR_3}, representing NBSRR, suggest that this design performs well only in the vicinity of the resonance, where its performance closely approaches the bound (electric size~$ka = 0.44$). At other electrical sizes, the design should be retuned to resonate at those sizes. The comparison of the performance of the realized designs and fundamental bounds can thus be simplified, making only the top projection of Fig.~\ref{fig:polarNBSRR_3}, as can be seen in Fig.~\ref{fig:polarNBSRR_2}. This representation is helpful for quantitatively assessing the performance. As an example, assume we desire to create a particle at electric size~$ka = 0.44$ with real part of normalized polarizability~$\T{Re} \left\{ \alpha_{zz}^\T{mm} \right\} / V = -10$. The fundamental bound in~Fig.~\ref{fig:polarNBSRR_2} tells us that the loss represented by the imaginary part of polarizability must be~$2 <  \T{Im} \left\{ \alpha_{zz}^\T{mm} \right\} / V < 50$, the lower bound being essential for creating low-loss metamaterial and the upper bound being essential when aiming for high extinction. Similarly, allowing at the same electric size only losses~$ \T{Im} \left\{ \alpha_{zz}^\T{mm} \right\} / V < 10$ means that the allowed range in the real part is approximately~$ -20 < \T{Re} \left\{ \alpha_{zz}^\T{mm} \right\} / V < 20$ and no design in the same region and using the same material can achieve wider range. 

\section{Generalization}
\label{sec:general}

\begin{figure}
    \centering
    \includegraphics[width=\linewidth]{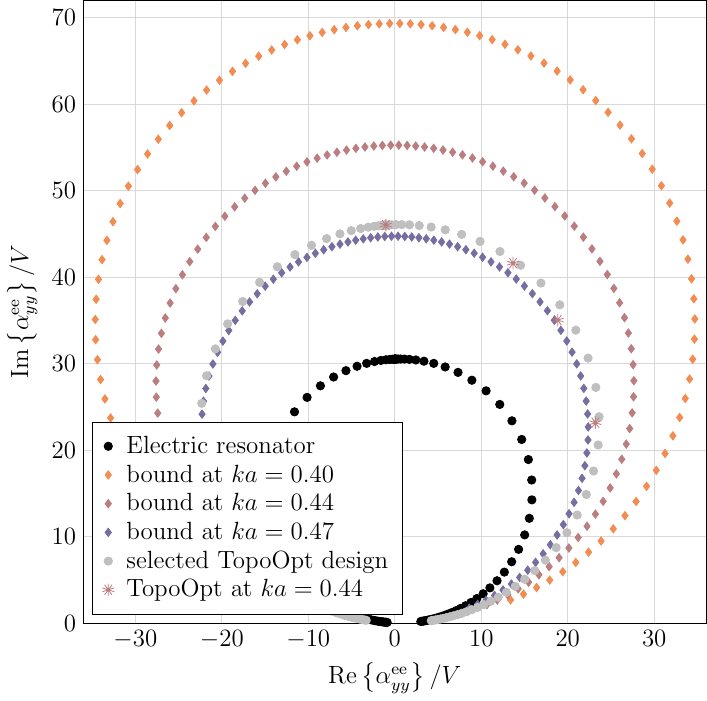}
    \caption{Normalized electric polarizability of an electric resonator depicted in Fig.~\ref{fig:FigNBSRR}c. The resonator is formed by carving a slot into the disc. Part of it is a circular slot of width~$2a/33$, and another part is a slot along the~$x$-axis of width~$a/33$. The bridges connecting the external and internal parts of the resonator have a width equal to the circular slot. The figure also shows the fundamental bound evaluated via~\eqref{eq:opt-problem1} at electric sizes~$ka = \left\{0.40, 0.44, 0.47 \right\}$. Results of topology optimization are shown by gray circles (frequency sweep of selected design) and star markers (best performance at $ka = 0.44$).}
    \label{fig:polarElectric_2}
\end{figure}

The modification of the fundamental bound to diagonal electric polarizability is straightforward. To this point, we can once more assume a disc region from Fig.~\ref{fig:FigNBSRR}a made of a good conductor. Also assume that high electric polarizability~$\alpha_{yy}^\T{ee}$ with vanishing magneto-electric coupling and vanishing electric cross-polarization is desired. An electric resonator proposed in~\cite{Schurig2006} and depicted in Fig.~\ref{fig:FigNBSRR}c could be a means to the desired goal. The particle resonates at approximately the same electric size as NBSRR from Sec.~\ref{sec:Theory} and its polarizability is depicted in Fig.~\ref{fig:polarElectric_2} in the same frequency band. The magneto-electric coupling is forbidden by inversion symmetry~\cite{LandauLifshitzPitaevskii1984}. The question is how well this scatterer performs and this is answered by comparing its performance to the fundamental bound formed by the optimization problem
\begin{equation}
    \begin{aligned}
\max \limits_\M{I} \quad & \left| p_y \right|^2  \\
  \T{s.t.} \quad     & \M{I}^\T{H} \M{ZI} = \M{I}^\T{H} \M{V} \\
  & \dfrac{ \T{Im} \left\{ \M{I}^\T{H} \M{V} \right\} } { \T{Re} \left\{ \M{I}^\T{H} \M{V} \right\} } = \beta , \,\beta  \in \left( - \infty ,\infty \right) \\
  & \V{m} = \M{0}, \, p_x = 0, \, p_z = 0
    \end{aligned}
    \label{eq:opt-problem1}
\end{equation}
which is analogous to~\eqref{eq:opt-problem}, but now assumes a sole excitation by electric field~$\V{E}^\T{inc} = E_y^\T{inc} \left(x,z \right) \M{y}_0$ that is approximately
constant around the origin~$E_y^\T{inc} \left( \M{0} \right) \approx 1 \, \T{Vm}^{-1} $. The last constraint enforces vanishing magneto-electric coupling and no cross-polarization. Using reasoning analogous to Sec.~\ref{sec:Theory}, we can claim that if a discoidal region of the same radius as the resonator and occupied by the same material is used, then no carving into the disc can result in higher electric polarizability~$\alpha_{yy}^\T{ee}$ than that achieved by the fundamental bound~\eqref{eq:opt-problem1}.

The fundamental bound for three electric sizes is depicted in~Fig.~\ref{fig:polarElectric_2}. In contrast with the magnetic polarizability discussed in Sec.~\ref{sec:IIA}, the performance of the realized design is considerably lower than the bound. This suggests that the design of the electric resonator depicted in Fig.~\ref{fig:FigNBSRR}c could likely be improved.

\begin{figure}
    \centering
    \includegraphics[width=0.8\linewidth]{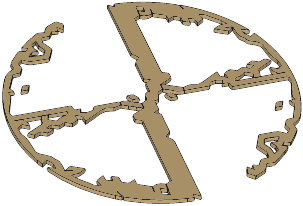}
    \caption{Selected design of an electrical resonator resulting from topology optimization. Its performance is depicted by gray circles in Fig.~\ref{fig:polarElectric_2}. The coordinate axes are defined in the same way as in Fig.~\ref{fig:FigNBSRR}a.}
    \label{fig:FigTopoOptElRes}
\end{figure}

Improvement can be achieved by topology optimization (TopoOpt). In this paper, the algorithm~\cite{Kadlec2026} running over the memetic scheme~\cite{Capek2023} is employed for its direct use of the electric-field integral equation. The design algorithm removes the metal from the disc region in an attempt to solve a problem analogous to~\eqref{eq:opt-problem1}, but the exact coupling between the material distribution and excitation via~\eqref{eq:Maxwell} is used instead of the power constraint~\eqref{eq:PextX1}. Several designs found by the algorithm at electric size~$ka = 0.44$ are shown in Fig.~\ref{fig:polarElectric_2} as star markers for different ratios of the real to imaginary parts of the electric polarizability. The performance of one of the designs (see Fig.~\ref{fig:FigTopoOptElRes} for geometry) is also plotted in Fig.~\ref{fig:polarElectric_2} as a sweep over electrical size (gray circles) to directly compare it to the original design (black circles). While the topmost point of the black trajectory (original resonator) achieves approximately~55~\% of the fundamental bound, the performance of the selected TopoOpt design reaches approximately~83~\% of the bound, which is a significant improvement.

\begin{figure}
    \centering
    \includegraphics[width=0.8\linewidth]{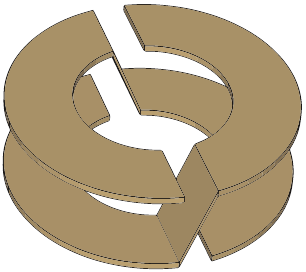}
    \caption{Chiral resonator with balanced polarizability components developed in~\cite{Marques2007}. Its performance is depicted by the black circles in Fig.~\ref{fig:ChSRREMZZ}. The coordinate axes are defined in the same way as in Fig.~\ref{fig:FigNBSRR}a. The external radius of each ring is~$r_\T{ext}$, the radius of the smallest circumscribing sphere is~$a = r_\T{ext} \sqrt{1469}/37$, the height of the resonator is equal to~$20r_\T{ext}/37$, the width of each ring is equal to $17r_\T{ext}/37$, and the ring split has a size equal to one half of the width.}
    \label{fig:FigchSRR}
\end{figure}

In the case of general polarizability tensors, the fundamental bound analogous to those presented above can also be formulated. Particularly, any polarizability component or their specific combination can be maximized in its absolute value (the first line of the optimization problem~\eqref{eq:opt-problem1}) by an appropriate combination of the electric/magnetic dipole moment and electric/magnetic type of excitation. The excitation is enforced by the power constraint, \textit{i.e.}, the second line of the optimization problem~\eqref{eq:opt-problem1}. On the other hand, the sweep over the polarizability phase, the third line of the optimization problem~\eqref{eq:opt-problem1}, represents a complicated combination of various polarizability components and the excitation fields, since
\begin{equation}
\M{I}^\T{H} \M{V} \approx \J \omega \varepsilon
    \mqty (\V{E}^\T{inc}\left( \M{0} \right) \\ c \V{B}^\T{inc} \left( \M{0} \right) )^\T{H}
    \mqty(\V{\alpha}_\T{ee} & \V{\alpha}_\T{em} \\ \V{\alpha}_\T{me} & \V{\alpha}_\T{mm} )^\T{H}
\mqty (\V{E}^\T{inc}\left( \M{0} \right) \\ c \V{B}^\T{inc} \left( \M{0} \right) ).
\label{eq:poweBalGen}
\end{equation}
To follow the reasoning used in the previous examples and to formulate bound, specific constraints on mutual polarizability dependencies are necessary. The construction of these constraints is demonstrated here in the example of a chiral particle presented in~\cite{Marques2007}, see Fig.~\ref{fig:FigchSRR}, which exhibits resonant polarizabilities~$\alpha_{zz}^\T{mm}, \alpha_{zz}^\T{ee}, \alpha_{zz}^\T{em} = - \alpha_{zz}^\T{me}$. The chiral particle is designed to be balanced with~$\alpha_{zz}^\T{mm} \approx \alpha_{zz}^\T{ee} \approx \pm \J \alpha_{zz}^\T{em}$.

Assume we would like to maximize the magneto-electric response of this particle described via polarizability component~$\alpha_{zz}^\T{em}$, keeping in mind that the particle should be balanced. For this we can choose a sole excitation by magnetic field~$\V{B}^\T{inc} = B_z^\T{inc} \left( x,y \right) \M{z}_0$ with~$c B_z^\T{inc} \left( \M{0} \right) = 1 \, \T{Vm}^{-1} $ and the maximization of the magnitude of the electric dipole moment component~$\left|p_z\right|$. Under sole magnetic excitation, the power balance~\eqref{eq:poweBalGen} gives
\begin{equation}
    \M{I}^\T{H} \M{V} \approx \J \omega m_z^* B^\T{inc}_z = \J \omega \varepsilon \alpha_{zz}^{\T{mm},*} \left| c B_z^\T{inc} \right|^2,
\end{equation}
which is not directly related to the desired polarizability component. This is resolved by an aditional constraint~$- \J c p_z = m_z$, which implies~$- \J \alpha_{zz}^\T{em} = \alpha_{zz}^\T{mm}  $  and
\begin{equation}
    \dfrac{\T{Re}\left\{ \M{I}^\T{H} \M{V}\right\}}{\T{Im}\left\{ \M{I}^\T{H} \M{V}\right\}} \approx 
    \dfrac{\T{Im}\left\{ \alpha_{zz}^\T{mm} \right\}}{\T{Re}\left\{ \alpha_{zz}^\T{mm} \right\}}
    =
    - \dfrac{\T{Re}\left\{\alpha_{zz}^\T{em} \right\}}{\T{Im}\left\{\alpha_{zz}^\T{em} \right\}}.
\end{equation}

The desired optimization problem, therefore, reads
\begin{equation}
    \begin{aligned}
\max \limits_\M{I} \quad & \left| p_z \right|^2  \\
  \T{s.t.} \quad     & \M{I}^\T{H} \M{ZI} = \M{I}^\T{H} \M{V} \\
  & \dfrac{ \T{Im} \left\{ \M{I}^\T{H} \M{V} \right\} } { \T{Re} \left\{ \M{I}^\T{H} \M{V} \right\} } = \beta , \,\beta  \in \left( - \infty ,\infty \right) \\
  & m_x = 0, \, m_y = 0, \, p_x = 0, \, p_y = 0 \\
  &\J c p_z = - m_z
    \end{aligned}
    \label{eq:opt-problem3}
\end{equation}
where we stress that the last two lines of constraints are affine constraints and, therefore, add no complexity to the solution, having been enforced prior to it~\cite{Liska_etal_FundamentalBoundsEvaluation}. This is true regardless of the number of affine constraints.

\begin{figure}
    \centering
    \includegraphics[width=\linewidth]{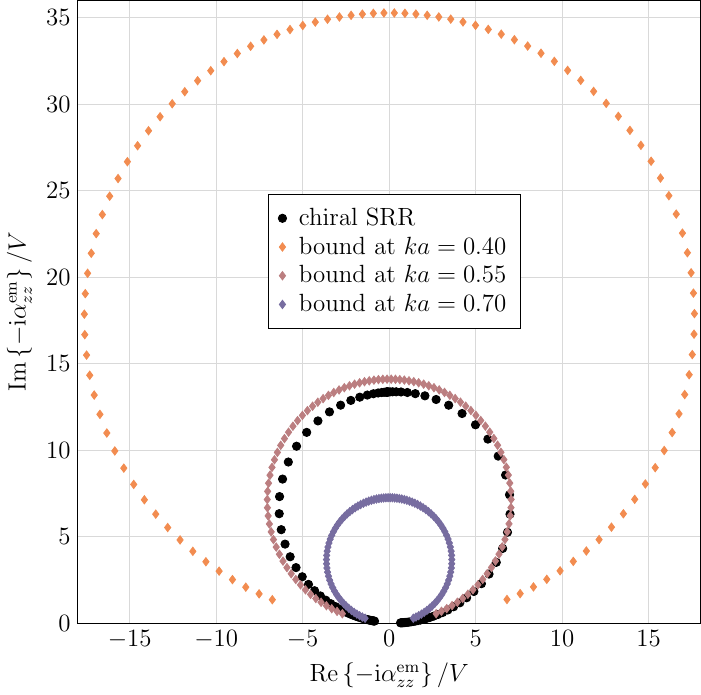}
    \caption{Normalized magneto-electric polarizability of a balanced chiral resonator from Fig.~\ref{fig:FigchSRR}. The resonant frequency is at electric size~$ka \approx 0.55$. The figure also shows the fundamental bound evaluated via~\eqref{eq:opt-problem3} at electric sizes~$ka = \left\{0.40, 0.55, 0.70 \right\}$.}
    \label{fig:ChSRREMZZ}
\end{figure}

\begin{figure}
    \centering
    \includegraphics[width=\linewidth]{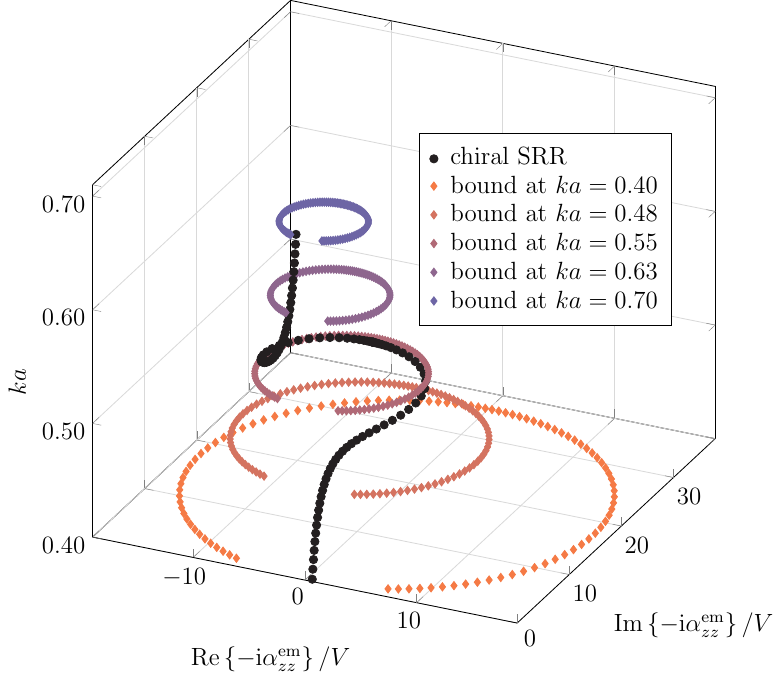}
    \caption{The polarizability data from~Fig.~\ref{fig:ChSRREMZZ} unwrapped along the frequency axis.}
    \label{fig:ChSRREMZZ1}
\end{figure}

The polarizability results are depicted in Fig.~\ref{fig:ChSRREMZZ}. The support of the optimal current density is chosen as a cylindrical surface concentric with the rings of the original resonator and tightly circumscribing it. As far as the balance between the polarizability components ($- \J \alpha_{zz}^\T{em} = \alpha_{zz}^\T{mm}  $) is required, the performance of the chiral resonator is nearly optimal in the vicinity of the resonance as seen by comparing the corresponding diamong markers and black circle markers in Fig.~\ref{fig:ChSRREMZZ} in their north region. The crossing of the markers representing the bound and the markers representing the design is merely a visual artifact of the two-dimensional projection of an otherwise three-dimensional black trajectory. The fundamental bound cannot be exceeded by any real design, which can be appreciated in the three-dimensional plot in Fig.~\ref{fig:ChSRREMZZ1}.

With respect to the analysis above, it is important to mention that the fundamental bound~\eqref{eq:opt-problem3} does not account for electric polarizability~$\alpha_{zz}^\T{ee}$ which is due to the use of a sole magnetic excitation. The fundamental bound with~$- \J \alpha_{zz}^\T{em} = \alpha_{zz}^\T{ee}  $ would form a separate optimization problem analogous to~\eqref{eq:opt-problem3} where magnetic and electric moments are interchanged and magnetic excitation is exchanged for electric excitation.






\section{Discussion}

This section summarizes the key takeaways from the previous sections and offers further guidance on using the developed fundamental bounds.

\subsection{Tightness of the Bounds}

The numerical results suggest that the bounds developed in this paper are relatively tight, meaning that a design closely approaching the bounding performance can be made. This is important for when the performance of a given design falls significantly behind the bound, it is highly likely that the design can be improved. The improved design might arise from topology optimization, as in Sec.~\ref{sec:general}, and the fundamental bound shows the value the design procedure should aim for.

\subsection{Studying Relations of Polarizability Components}

\begin{figure}
    \centering
    \includegraphics[width=\linewidth]{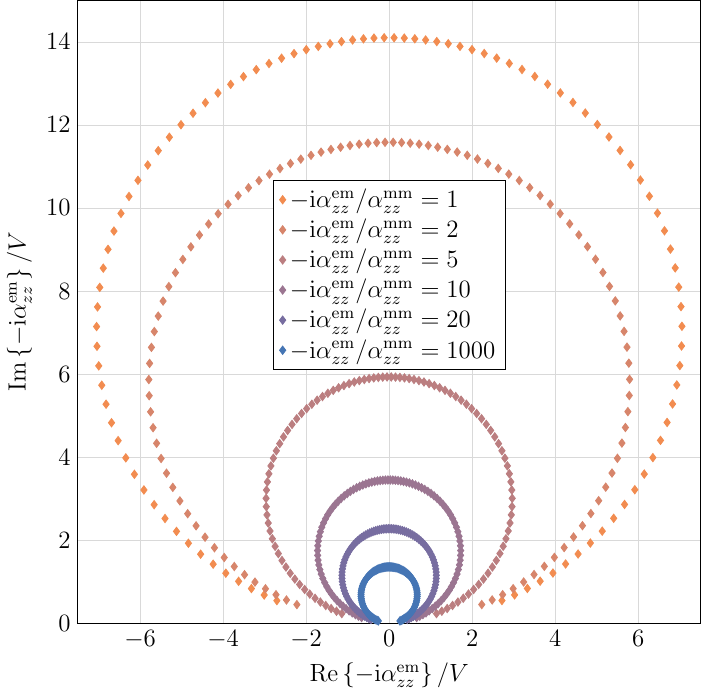}
    \caption{Fundamental bound on the normalized magneto-electric polarizability of a cylindrical current support used in Fig.~\ref{fig:ChSRREMZZ}. The electric size~$ka \approx 0.55$. The fundamental bound was evaluated via~\eqref{eq:opt-problem3} but with different ratios between magnetic-magnetic and electric-magnetic polarizability components. The largest response belongs to ratio~$- \T{i} \alpha^\T{em}_{zz} / \alpha^\T{mm}_{zz} = 1$ and monotonically decays to the smallest response for ratio~$- \T{i} \alpha^\T{em}_{zz} / \alpha^\T{mm}_{zz} = 1000$.}
    \label{fig:PronouncedEM}
\end{figure}

The fundamental bounds can be used to study inherent coupling between different polarizability components. This idea is presented in Fig.~\ref{fig:PronouncedEM} which shows a fundamental bound~\eqref{eq:opt-problem3} on the same current support and the same material but with the last constraint in the form~$- \J c p_z = x m_z$ (implying~$- \J \alpha_{zz}^\T{em} = x \alpha_{zz}^\T{mm}  $) where parameter~$x$ fixes the relative strength of magnetic-magnetic and electric-magnetic components. We might, for example, ask what is the cost of creating a scatterer that would mostly exhibit electric-magnetic polarizability while having a negligible magnetic-magnetic component~\cite{Albooyeh2016}, \textit{i.e.}, a scatterer with a high value of parameter~$x$. The result is shown in Fig.~\ref{fig:PronouncedEM}. It can clearly be observed that the cost of enhancing the electro-magnetic coupling is dire. A scatterer with an almost pure electro-magnetic response can possibly be realized but the strength of its response would be minuscule. All this can be stated without knowledge of the specific geometry of such a scatterer.

\subsection{Different Current Supporting Regions And Different Materials}

The fundamental bound provides an absolute measure of the performance of the given current-supporting region and material, all without requiring knowledge of the specific designs that achieve this performance. For example, when studying the magnetic response of NBSRR, the fundamental bound has been evaluated for a disc of the same material that tightly circumscribes the resonator. The performance of NBSRR was very close to the bound, with little room for improvement. We can, however, ask if the performance could be improved by extending the planar design into the third dimension. What if the resonator had been carved into the surface of a highly conducting sphere of the same electrical size? The design of the resonator might not be clear but by evaluating the fundamental bound using the same optimization problem~\eqref{eq:opt-problem}, we would find that, with a sphere as the design region, the maximum improvement over a disk is approximately 7~\%, which is rather small. It is worth noting that in the case of a highly conducting spherical surface, the top-most point in the plot like in~Fig.~\ref{fig:polarNBSRR_2}, which corresponds to~$\T{Re} \left\{\alpha^\T{mm}_{zz} \right\} / V = 0$, can be evaluated analytically as
\begin{equation}
   \dfrac{\T{Im} \left\{\alpha^\T{mm}_{zz} \right\}}{V}
     = \dfrac{3}{2} \dfrac{\T{j}_1 \left(ka \right)}{ \left( ka \right)^2 \T{j}_1^2 \left(ka \right) + \dfrac{Z_s}{Z_0}},
\end{equation}
where~$Z_s$ is the surface impadance of the conductor,~$Z_0$ is the free-space impedance and~$\T{j}_1$ is the spherical Bessel function of the first order. The derivation stems from using vector spherical waves~\cite{Kristensson_ScatteringBook} as basis functions~$\V{\psi}_i$ from Appendix~\ref{app:A} expanding the optimal current density.

\begin{figure}
    \centering
    \includegraphics[width=0.8\linewidth]{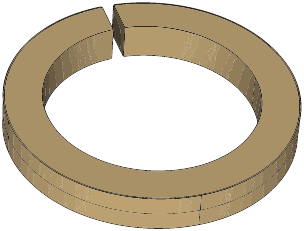}
    \caption{Silver split ring resonator studied in~\cite{Delgado2009}. Its performance is depicted by the black circles in Fig.~\ref{fig:AgSRRMMZZ}. The coordinate axes are defined in the same way as in Fig.~\ref{fig:FigNBSRR}a.}
    \label{fig:FigAgSRR}
\end{figure}
\begin{figure}
    \centering
    \includegraphics[width=\linewidth]{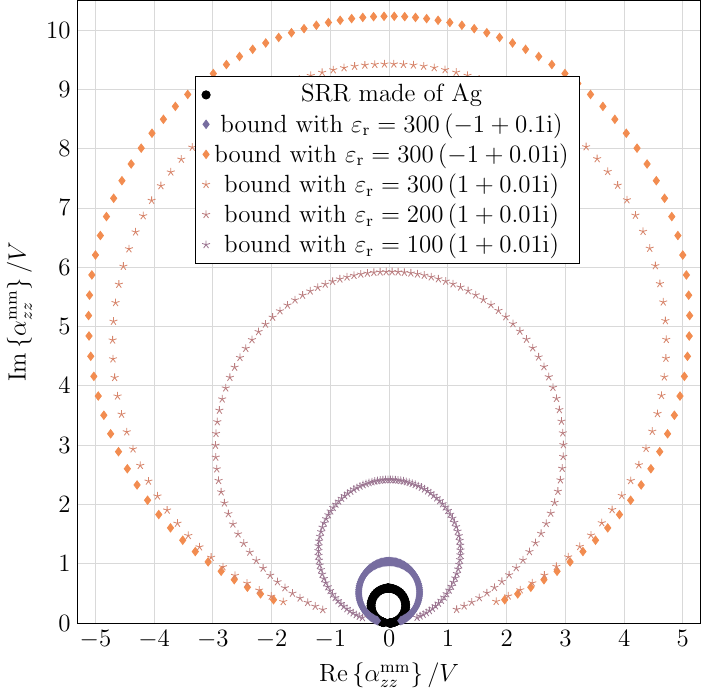}
    \caption{Normalized magnetic polarizability of a split ring resonator made of silver, see Fig.~\ref{fig:FigAgSRR} for the geometry. The orientation of the resonator is the same as that of NBSRR in Fig.~\ref{fig:FigNBSRR}b. The external radius is~$r_\T{ext} = 50 \, \T{nm}$, the height of the resonator is equal to~$6r_\T{ext}/23$, the width of the ring is equal to the height, and the ring split has a size equal to one third of the height. The resonant frequency is at electric size~$ka \approx 0.12$ and corresponds to wavelength~$\lambda \approx 2600 \, \T{nm}$. The figure also shows the fundamental bound evaluated via~\eqref{eq:opt-problem} at an electric size~$ka \approx 0.12$ and for different materials filling the current-supporting region (a cylindrical volume tightly circumscribing the original resonator). The relative permittivity~$\varepsilon_\T{r} = 300 \left(-1 + 0.1 \J \right)$ corresponds to silver at wavelength~$\lambda \approx 2600 \, \T{nm}$.}
    \label{fig:AgSRRMMZZ}
\end{figure}

In a similar manner, we might ask about the suitability of the material used. The NBSRR studied in Sec.~\ref{sec:Theory} was made of copper at microwave frequencies, \textit{i.e.}, from an excellent, low-loss conductor. Its magnetic response was high and close to the fundamental bound. Quite a different result can be found at optical frequencies, as shown in Fig.~\ref{fig:AgSRRMMZZ} which presents a magnetic polarizability~$\alpha_{zz}^\T{mm}$ of a split-ring resonator described in~\cite{Delgado2009}, the geometry of which is shown in Fig.~\ref{fig:FigAgSRR} and which is made of silver. The polarizability exhibits resonance at wavelegth~$\lambda \approx 2600 \, \T{nm}$ at which the silver can be described by relative permittivity~$\varepsilon_\T{r} = 300 \left(-1 + 0.1 \J \right)$ with significant loss tangent. The loss is the main reason for the extremely low polarizability response represented by the black circle markers, and even the fundamental bound evaluated at the resonant wavelength shows that the response cannot be made much higher. Comparing this result to the performance of NBSRR from Sec.~\ref{sec:Theory}, the response of the silver ring is approximately 50~times weaker. While in the case of copper at microwave frequencies, the response was limited by radiation, in the case of silver at optical frequencies, the loss dominates the extinct power. The fundamental bound tells us that a substantial improvement cannot be achieved by geometry. The only possibility is to change the material. This is clearly presented by other markers in Fig.~\ref{fig:AgSRRMMZZ} which show that a large magnetic response is only achievable when losses are small and the dielectric contrast is high. A viable option is, therefore, to use low-loss dielectric materials with a high dielectric constant, an idea studied in~\cite{Popa2008,Jelinek2009}.

\section{Conclusion}

The fundamental bounds on components of polarizability tensors provide an absolute measure of their maximal strength. They have been developed into a versatile framework that allows tackling a wide range of practical scenarios. Several examples demonstrated how to work with the developed framework. In particular, it has been shown how to isolate polarizability components via additional linear constraints, or how to use bounds to assess the polarizability potential of a given design region and/or the material from which the polarizable scatterer is to be built. Intuitive visualization has been suggested, enabling an easy comparison between the performance of realized designs and their corresponding upper bounds.

By comparing the fundamental bounds with the polarizability of known designs at the microwave and plasmonic frequency ranges, it has been shown that split-ring resonators made of good conductors are close to optimal in their magnetic response, while the performance of known electric resonator designs falls significantly short of their potential. More specifically, the design obtained via topology optimization improved the electrical response by more than 50\%. Analogously, the bounds have been used to demonstrate the high cost of using plasmonic metals to create artificial magnetism or the high cost of boosting magneto-electric response.

Thanks to the developed framework, the polarizability potential of existing and future designs of polarizable particles can be unambiguously quantified, and such studies might reveal a performance gap for existing designs, opening a possibility to improve the performance of devices based on them.

\begin{acknowledgments}
This work was supported by the Czech Science Foundation under project~\mbox{No.~24-11678S}.
\end{acknowledgments}

\appendix

\section{Electric Field Integral Equation}
\label{app:A}

Assuming a time-harmonic steady state at angular frequency~$\omega$ and an electric field~$\V{E}^\T{inc}$ incident on an electrically polarizable~\footnote{Magnetized matter is omitted in this paper.} obstacle in an otherwise empty space, the solution to Maxwell's equations can be, using volume equivalence~\cite{Harrington2001}, reformulated into
\begin{equation}
    \V{E}^\T{sca} \left\{ \V{J} \left( \V{r} \right) \right\} + \V{E}^\T{inc} \left( \V{r} \right) = \V{\rho} \left( \V{r} \right) \cdot \V{J} \left( \V{r} \right), 
    \label{eq:EFIEoper}
\end{equation}
where~$\V{E}^\T{sca}$ is an operator transforming free-space current density~$\V{J}$ into a scattered electric field and~$\V{\rho}$ is the resistivity tensor, defined via material relation 
\begin{equation}
\V{J} \left( \V{r} \right) = - \T{i} \omega \varepsilon_0 \V{\chi} \left( \V{r} \right) \cdot \V{E} \left( \V{r} \right) = \V{\rho}^{-1} \left( \V{r} \right) \V{E} \left( \V{r} \right),
\end{equation}
with~$\varepsilon_0$ being the vacuum permittivity and~$\V{\chi}$ being the dimensionless electric susceptibility tensor.

Relation~\eqref{eq:EFIEoper} is an integro-diferential equation for unknown polarization current density~$\V{J}$ which is typically approached via Galerkin's method~\cite{Kantorovich1982,Harrington1993} which represents the current density using a set of basis functions~$\left\{ \V{\psi}_i \right\}$
\begin{equation}
    \V{J} \left( \V{r} \right) \approx \sum \limits_{i} I_i \V{\psi}_i \left( \V{r} \right),
    \label{Eq:CurrExp}
\end{equation}
and transforms the linear operator equation~\eqref{eq:EFIEoper} into a system of linear equations
\begin{equation}
    \left( \M{Z}_0 + \M{Z}_\rho \right) \M{I} = \M{V},
    \label{eq:EFIEmat}
\end{equation}
where
\begin{equation}
    z^0_{mn} = - \langle \V{\psi}_m^*,  \V{E}^\T{sca} \left\{ \V{\psi}_n  \right\} \rangle ,
    \label{eq:z0mn}
\end{equation}
are elements of free-space impedance matrix~$\M{Z}_0$, 
\begin{equation}
    z^\rho_{mn} = \langle \V{\psi}_m^*,  \V{\rho} \cdot \V{\psi}_n  \rangle
    \label{eq:zrhomn}
\end{equation}
are elements of material impedance matrix~$\M{Z}_\rho$ and
\begin{equation}
    v_{m} = \langle \V{\psi}_m^*,  \V{E}^\T{inc}  \rangle
    \label{eq:vm}
\end{equation}
are elements of the excitation vector. In the above, symbols
\begin{equation}
    \langle \V{a},  \V{b}  \rangle = \int \limits_{V} \V{a}^* \cdot \V{b} \, \T{d} V
    \label{eq:sProd}
\end{equation}
denote scalar product with~$^*$ being complex conjugation.

In the context of Galerkin's method, the electric and magnetic dipole moments
\begin{equation}
    \V{p} = - \dfrac{1}{ \J \omega} \int\limits_V \V{J} \T{d} V, \quad {\V{m}} = \dfrac{1}{2}\int\limits_V {\V{r} \times {\V{J}}\T{d} V}
    \label{eq:mpDef}
\end{equation}
become linear forms of vector~$\M{I}$, namely
\begin{equation}
    \V{p} \cdot \V{v} = \M{P}^v \M{I}, \quad \V{m} \cdot \V{v} = \M{M}^v \M{I},
\end{equation}
where
\begin{equation}
P^v_m = - \dfrac{1}{ \J \omega} \langle \V{\psi}_m^*,  \V{v}  \rangle, \, M^v_m = \dfrac{1}{2} \langle \V{r} \times \V{\psi}_m^*,  \V{v}  \rangle    
\end{equation}
with~$\V{v}$ being an arbitrary constant vector.

\section{Expansion of the Extinct Power}
\label{app:B}

Assume integral
\begin{equation}
    \int\limits_V \V{J}^* \cdot \V{E}^\T{inc}\T{d} V  = \M{I}^\T{H} \M{V}.
    \label{eq:Pext}
\end{equation}
Assume further the extent of the sources is small enough that incident field~$\V{E}^\T{inc}$ is well represented by the first two terms of its Taylor series
\begin{equation}
    \V{E}^\T{inc}\left( \V{r} \right) = \V{E}^\T{inc} \left( \M{0} \right) + \sum \limits_\alpha  {r_\alpha \partial _\alpha \V{E}^\T{inc} \left( \M{0} \right)} + \dots
    \label{eq:Taylor}
\end{equation}
where~$r_\alpha$ represents the components of the radius vector and the argument~$\left( \M{0} \right)$ signifies that the derivatives are evaluated at the origin which coincides with the center of the smallest sphere circumscribing the sources.

Substituting~\eqref{eq:Taylor} into~\eqref{eq:Pext} gives
\begin{equation}
    \M{I}^\T{H} \M{V} =  \J \omega \V{p}^* \cdot \V{E}^\T{inc} \left( \M{0} \right) + \sum \limits_{\alpha ,\beta } \partial_\beta E_\alpha^\T{inc} \left( \M{0} \right) \int\limits_V r_\beta J_\alpha^* \T{d} V + \dots
    \label{eq:AppB1}
\end{equation}
with~$\V{p}$ being the electric dipole moment defined by~\eqref{eq:mpDef}.

The summation term in~\eqref{eq:AppB1} can be written as
\begin{multline}
  \sum \limits_{\alpha ,\beta } \partial_\beta E_\alpha ^\T{inc} \left( \M{0} \right) \int\limits_V r_\beta J_\alpha^* \T{d} V  = \\ \dfrac{1}{4} \sum\limits_{\alpha ,\beta } \left( \partial_\beta E_\alpha^\T{inc} \left( \M{0} \right) - \partial_\alpha E_\beta ^\T{inc} \left( \M{0} \right) \right) \int\limits_V \left( r_\beta J_\alpha^* - r_\alpha J_\beta^* \right) \T{d} V \\
   + \dfrac{1}{2} \sum\limits_{\alpha ,\beta } \partial_\beta E_\alpha^\T{inc} \left( \M{0} \right) \int\limits_V \left( r_\beta J_\alpha^* + r_\alpha J_\beta^* \right) \T{d} V.
   \label{eq:AppB2}
\end{multline}

The first term in the RHS of~\eqref{eq:AppB2} is recognized as the curl of the electric field at the origin in scalar multiplication with the magnetic dipole moment~$\V{m}$ defined by~\eqref{eq:mpDef}. The second term relates the electric field gradient to the quadrupole tensor. 

\begin{figure}
    \centering
    \includegraphics[width=\linewidth]{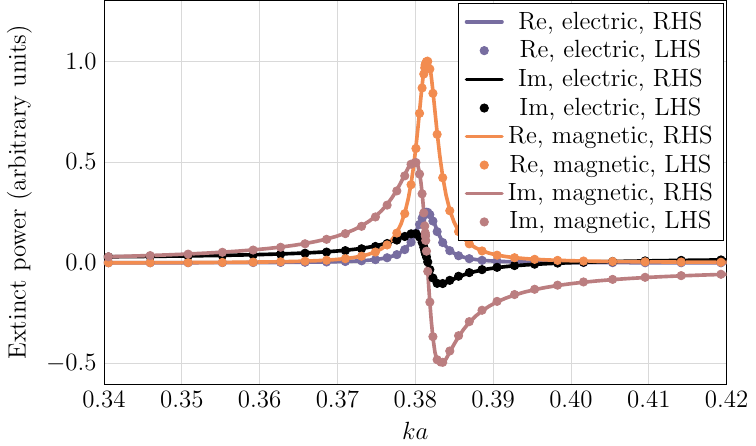}
    \caption{The curves and markers show the left- and right-hand sides of~\eqref{eq:PextApprox} for an edge-coupled split-ring resonator under two dual excitations. The excitation (electric/magnetic) is labeled in the legend of the figure. The external radius of the resonator is~$a$. The width of the strips is~$5a /28$, the width of the slot is~$a/28$, and the width of the ring splits is~$a/14$.}
    \label{fig:Pext}
\end{figure}

Summarizing, there is 
\begin{equation}
    \M{I}^\T{H} \M{V} =  \J \omega \V{p}^* \cdot \V{E}^\T{inc}\left( \M{0} \right) + \J \omega \V{m}^* \cdot \V{B}^\T{inc} \left( \M{0} \right) + \dots,
    \label{eq:PextApprox}
\end{equation}
where Faraday’s law
\begin{equation}
    \nabla  \times \V{E}^\T{inc} \left( \M{0} \right) =  \J \omega \V{B}^\T{inc} \left( \M{0} \right)
\end{equation}
was employed, and the quadrupole and higher terms are not shown (are neglected). In typical conditions, the omission of these terms is acceptable, as presented in~Fig.~\ref{fig:Pext} on an example of an edge-coupled split ring resonator (ECSRR)~\cite{PendryHoldenRobbinsEtAl1999}. The resonator is centered at the origin with its axis pointing along direction~$\M{z}_0$, similarly to Fig.~\ref{fig:FigNBSRR}a. The ring splits are positioned on the~$x$-axis. The excitation of such a resonator via a homogeneous electric field (vanishing magnetic field) along the~$y$-axis results in a resonant electric dipole moment along the same direction and a resonant magnetic dipole moment along the~$z$-axis. The excitation of such a resonator via a homogeneous magnetic field (vanishing electric field) along the~$z$-axis results in a resonant magnetic dipole moment along the same direction and a resonant electric dipole moment along the~$y$-axis. One possibility of how to realize such excitation is given in Appendix~\ref{app:C}. Despite the complex electromagnetic behavior of this resonator (resonant electric and magnetic polarizability and significant bianisotropy), the approximation~\eqref{eq:PextApprox} is excellent.

\section{Particle Excitation}
\label{app:C}

The excitation of electrically small scatterers in the context of their polarizability is commonly assumed to be uniform. Since this paper deals with full dynamic solutions and uses the electric-field integral equation, which assumes the incident field satisfies free-space Maxwell's equations, this assumption must be relaxed. The uniform excitation is, in this paper, substituted by cylindrical waves assuming that the smallest sphere of radius~$a$ circumscribing the scatterer is electrically small, \textit{i.e.},~$ka \ll 1$. 

For example, demanding a uniform electric field and vanishing magnetic field around the origin, the excitation is taken as
\begin{equation}
\begin{aligned}
\V{E}^\T{inc} \left( \V{r} \right) &= {{\M{z}}_0}{\T{J}_0}\left( {k \sqrt {{x^2} + {y^2}} } \right),\\
{c}\V{B}^\T{inc} \left( \V{r} \right) &= \J \dfrac{{ {{\M{x}}_0}y - {{\M{y}}_0}x}}{{\sqrt {{x^2} + {y^2}} }} {\T{J}_1}\left( {k \sqrt {{x^2} + {y^2}} } \right),
\label{AppC:1}
\end{aligned}
\end{equation}
where $\T{J}_i$ represents Bessel's function of the first kind, and~$c$ is speed of light. Demanding a uniform magnetic field and vanishing electric field around the origin, the excitation is taken as
\begin{equation}
\begin{aligned}
\V{E}^\T{inc} \left( \V{r} \right) &=  - \J \dfrac{{ {{\M{x}}_0}y - {{\M{y}}_0}x}}{{\sqrt {{x^2} + {y^2}} }} {\T{J}_1}\left( {k \sqrt {{x^2} + {y^2}} } \right),\\
{c}\V{B}^\T{inc} \left( \V{r} \right) &= {{\M{z}}_0}{\T{J}_0}\left( {k \sqrt {{x^2} + {y^2}} } \right).
\label{AppC:2}
\end{aligned}
\end{equation}
Other field directions are achieved by rotating~\eqref{AppC:1} or~\eqref{AppC:2}.

\end{document}